\documentclass[twocolumn]{aastex63}
\usepackage{array}
\usepackage{multirow}
\usepackage{xcolor} 
\usepackage{amsmath}

\newcolumntype{L}{>{\centering\arraybackslash}m{3cm}}
\newcolumntype{M}{>{\centering\arraybackslash}m{1cm}}

\received{June 8, 2021}
\revised{August 2, 2021}
\accepted{August 5, 2021}

\submitjournal{ApJ}

\shorttitle{Scaling of Small-Scale Dynamo Properties in the Rayleigh-Taylor Instability}
\shortauthors{V. Skoutnev, E. R. Most, A. Bhattacharjee, A. A. Philippov}
\graphicspath{{./}{}}

\begin{document}

\title{Scaling of Small-scale Dynamo Properties in the Rayleigh-Taylor Instability}

\correspondingauthor{Valentin Skoutnev}
\email{skoutnev@princeton.edu}

\author{V. Skoutnev}
 \affiliation{Department of Astrophysical Sciences and Max Planck Princeton Center, Princeton University, Princeton, NJ 08544, USA}
\author{E. R. Most}
\affiliation{Princeton Center for Theoretical Science, Princeton University, Princeton, NJ 08544, USA}
\affiliation{Princeton Gravity Initiative, Princeton University, Princeton, NJ 08544, USA}
\affiliation{School of Natural Sciences, Institute for Advanced Study, Princeton, NJ 08540, USA}
\author{A. Bhattacharjee}
\affiliation{Department of Astrophysical Sciences and Max Planck Princeton Center, Princeton University, Princeton, NJ 08544, USA}
\author{A. A. Philippov}
\affiliation{Center for Computational Astrophysics, Flatiron Institute, 162 Fifth Avenue, New York, NY10010, USA}

\begin{abstract}
We derive scaling relations based on freefall and isotropy assumptions for the kinematic small-scale dynamo growth rate and amplification factor over the course of the mixing, saturation, and decay phases of the Rayleigh-Taylor instability (RTI) in a fully-ionized plasma. The scaling relations are tested using sets of three-dimensional, visco-resistive MHD simulations of the RTI. They are found to hold in the saturation phase, but exhibit discrepancies during the mixing and decay phases, suggesting a need to relax either the freefall or isotropy assumptions. Application of the scaling relations allows for quantitative prediction of the net amplification of magnetic energy in the kinematic dynamo phase and therefore a determination of whether the magnetic energy either remains sub-equipartition at all velocity scales or reaches equipartition with at least some scales of the turbulent kinetic energy in laboratory and astrophysical scenarios. As an example, we consider the dynamo in RTI-unstable regions of the outer envelope of a binary neutron star merger, and predict that the kinematic regime of the small-scale dynamo ends on the time scale of nanoseconds and then reaches saturation on a timescale of microseconds, which are both fast compared to the millisecond relaxation time of the post-merger.

\end{abstract}


\section{Introduction} \label{sec:intro}

The  Rayleigh-Taylor instability (RTI) is ubiquitous in astrophysical contexts due to the generality of the conditions needed for its onset. The RTI operates in regions where the density gradient is misaligned with the direction of the local gravitational field or acceleration \citep{Chandrasekhar1961}. The instability characteristically evolves with rising bubbles of lighter fluid and sinking spikes of heavier fluid that propagate away from the unstable region, leading to mixed fluids and relaxation of the unstable density gradient. In this manner, fluid mixing and transport is enhanced by the RTI in many astrophysical scenarios such as the solar corona \citep{Isobe2005,Berger2011,hillier2018magnetic}, gamma-ray burst scenarios  \citep{gull1973,Levinson_2009,Duffell2013,Duffell2014}, and supernova explosions \citep{HillebrandtNiemeyer,Cabot2006,Duffell2016}. The fluid is typically a highly conducting plasma whose ability to generate and sustain magnetic fields can make the magnetohydrodynamic (MHD) version of the RTI differ from its hydrodynamic counterpart. 

The role of the magnetic field can be categorized by whether the initial field strength is dynamically strong or weak. In the strong field limit, large scale magnetic fields are able to stabilize a wide range of wavenumbers due to the restoring force of magnetic tension \citep{Chandrasekhar1961,Ruderman_2014}. This affects the nonlinear saturation and mixing rates of the RTI with a strong dependence on details of the initial geometry of the magnetic fields. Typically, magnetic tension forces suppress development of secondary shear instabilities and can cause bubble/spike structures to rise/fall more rapidly than in the hydrodynamic case \citep{Stone2007}. This limit has application in the contexts of solar prominences \citep{hillier2018magnetic} and pulsar wind nebulae \citep{Porth2014}, for example.

On the other hand, the weak field limit corresponds to an initial magnetic field with dynamically insignificant strength, which allows the RTI to evolve purely hydrodynamically at least at early times. The turbulence in the nonlinear phases of the RTI has the potential to amplify the magnetic field through dynamo action. In the dynamo literature, this weak field limit is also known as the kinematic dynamo regime. If the turbulence is sufficiently vigorous and the initial magnetic energy is not too small, then the magnetic energy could grow to and saturate in equipartition with the kinetic energy of the turbulence. This results in decaying MHD turbulence post-saturation of the RTI. Otherwise, the magnetic energy is amplified, but remains at sub-equipartition energies and results in decaying hydrodynamic turbulence post-saturation of the RTI.

While early, low resolution simulations of the RTI had confirmed operation of the small-scale dynamo (SSD) \citep{Jun1995}, the quantitative scaling of dynamo properties has been largely unstudied despite applications in a variety of scenarios. Here, we discuss two example applications at opposite extremes: the fireball and laser plasma experiments. Hydrodynamical simulations of the fireball propose that the high-Reynolds-number RTI-driven turbulence generates equipartition magnetic fields that can explain the observed levels of synchrotron radiation emission from these sources \citep{Duffell2013}.  On the other hand, several laser plasma experiments of the RTI at modest Reynolds numbers have found magnetic energy growth and explain the observations by using the Biermann effect \citep{gao2012magnetic,manuel2012first,Nilson2015,JMatt_Biermann}. However, RTI turbulence itself could be a significant contributor as well, in particular as the Reynolds number and duration of future experiments increases \citep{Galmiche_1996,bott2021time}. A quantitative framework for either justifying the use of the equipartition argument or otherwise predicting the level of sub-equipartition magnetic energy amplification would be useful in the general case. This motivates a careful study of the RTI-turbulence-driven dynamo.

\paragraph{Paper Outline} Section \ref{sec:TheoreticalScalingAnalysis} presents a model for the SSD in the weak field limit based on assumptions of isotropic turbulence and freefall scaling for the outer velocity and length scale of the turbulence. The model makes predictions for scaling laws between the SSD growth rate, amplification factor, and parameters of the RTI (Atwood number, gravitational acceleration, viscosity, and length scale) in each phase of evolution of the RTI. Determination of the correct scaling laws is important for quantitatively extending results from practical simulation parameters to realistic experimental or astrophysical parameters, which can be separated by orders of magnitude. 

Section \ref{sec:SimulationScalingAnalysis} tests the model using sets of three-dimensional (3D) visco-resistive MHD direct numerical simulations that resolve the turbulent viscous scales, allowing for results independent of numerical resolution. Discrepancies between the model prediction and the numerical results are analyzed and the assumptions that are likely breaking down are identified. Section \ref{sec:Application} applies the model to the case of possible dynamo action in RTI-unstable regions of the outer envelope of the post-merger of a binary neutron star collision. In Section \ref{sec:Conclusion}, we summarize our findings, identify directions for future work, and conclude.

\section{Theoretical Scaling Predictions}\label{sec:TheoreticalScalingAnalysis}
We present scaling arguments to determine the kinematic small-scale dynamo growth rate and amplification factor of the magnetic energy in a Rayleigh-Taylor unstable, conducting, collisional fluid over the course of the growth, saturation, and decay of the instability. 

\paragraph{Setup} Without loss of generality, we consider the standard, idealized RT setup in three dimensions with a discontinuous density jump from $\rho(z)=\rho_b$ for $z<0$ to $\rho(z)=\rho_t>\rho_b$ for $z>0$ in a bounded domain of size $\sim L$ and initial velocity perturbations near $z=0$. Any similar setup (e.g. with a continuous positive density gradient instead) can be easily related to the idealized setup, and the same scaling arguments will apply. The initial seed magnetic field is taken to be random at all scales and arbitrarily weak so that the magnetic energy always remains much lower than the kinetic energy at all hydrodynamic scales, resulting in purely hydrodynamic evolution of the velocity field. A standard model of the RTI in fully ionized plasmas is the set of visco-resistive MHD equations given by:

\begin{equation}\label{RMHD1}
\partial_t \rho+ \nabla \cdot\left(\rho \textbf{u}\right)=0,
\end{equation}

\begin{equation}\label{RMHD2}
\partial_t (\rho \textbf{u})+\nabla \cdot \left(\rho\textbf{u}\textbf{u}-\textbf{B}\textbf{B}+P \right)=-\rho g\hat{z}+ \rho\nu\nabla^2\textbf{u},
\end{equation}

\begin{equation}\label{RMHD3}
\partial_t \textbf{B}-\nabla\times \left(\textbf{u}\times \textbf{B}\right)=\eta\nabla^2\textbf{B},
\end{equation}

\begin{equation}\label{RMHD4}
\partial_t E+\nabla \cdot \left[ (E+P)\textbf{u}-\textbf{B}(\textbf{B}\cdot\textbf{u})\right]=-\rho g u_z +\rho\kappa\nabla^2 T,
\end{equation}
where $\textbf{u}$ is the velocity, $\textbf{B}$ is the magnetic field,  $E$ is the total energy density, $P$ is the sum of the gas and magnetic pressure, $T$ is the temperature, $g$ is the gravitational acceleration, $\nu$ is the viscosity, $\eta$ is the resistivity, and $\kappa$ is the thermal diffusivity. 

We consider the subsonic limit where the fluid flows are nearly incompressible. Any initial vertical velocity perturbation with horizontal wavenumber $k$ triggers the RTI and grows exponentially with rate $n=(g k A+\nu^2k^4)^{\frac{1}{2}}-\nu k^2$,
where $A=\frac{\rho_t-\rho_b}{\rho_t+\rho_b}\leq1$ is the Atwood number and effects of thermal diffusivity are neglected. If all wavenumbers are present, the fastest growing wavenumber will be $k_c\sim (Ag/\nu^2)^{\frac{1}{3}}$ and smaller wavenumbers $k<k_c$ will approximately have the inviscid growth rate $n\approx (Agk)^{\frac{1}{2}}$. 

\paragraph{RTI evolution and definitions} The RTI can be split into four distinct phases: linear growth, mixing (or nonlinear), saturation, and decay. The linear phase begins with exponential growth of any initial vertical velocity perturbations and ends once the fluid has displaced a vertical distance comparable to the wavelength of the initial mode. The fluid then rapidly becomes turbulent with outer velocity scale $u(t)$, integral length scale $l_i(t)$, and Reynolds number $Re(t)=u(t)l_i(t)/(2\pi\nu)$ evolving in time throughout the remaining three phases. It is convenient in what follows to define a characteristic velocity $u_{sat}=\sqrt{AgL}$, dynamical time $t_{dyn}=L/u_{sat}$, and Reynolds number $Re_{sat}=u_{sat}L/(2\pi \nu)$.

Following the linear phase, the initial perturbations develop in the mixing phase into the characteristic RTI bubble/spike structures that rise/fall away from the boundary, driven by buoyancy forcing. The mixing phase ends when the upward/downward propagating fronts of bubbles/spikes reach the system scale $L$. This begins the saturation phase of the RTI where the majority of available potential energy has been released into turbulent kinetic energy. After a dynamical time scale, the viscous dissipation in the turbulence becomes larger than the energy input from buoyant forcing and thus begins the decay phase where the total kinetic energy decreases. The fluid eventually settles into a stable state with a negative density gradient.   

The small-scale dynamo will be active when the Reynolds number $Re(t)$ in the turbulent phases is above the critical Reynolds number $Re^c(Pm)$ of the flows, where $Pm=\nu/\eta$ is the magnetic Prandtl number.  We characterize the dynamo with the instantaneous, exponential dynamo growth rate $\gamma(t)=\frac{d}{dt}\ln ME(t)$ and the net exponential amplification factor of the magnetic energy given by:

\begin{equation}\label{MAF}
\Delta=\max_t\ln\left(\frac{ME(t)}{ME(0)}\right)=\max_t\int_0^{t}\gamma(t')dt',
\end{equation}
where $\mathrm{ME}(t)=\frac{1}{2}\int \textbf{B}(\textbf{x},t)^2d^3x$ is the total magnetic energy. If $u$, $l_i$, and $Re$ are enough to characterize the flow, then dimensional constraints force $\gamma(t)= C_\gamma (u/l_i) Re^{D_{\gamma}}$ (assuming a power-law form for the $Re$ dependence when $Re\gg Re^c$), where $C_\gamma$ and ${D_{\gamma}}$ are dimensionless constants. Since we consider the case of an arbitrarily weak initial magnetic energy, the dynamo always remains in the kinematic regime and the dynamo growth rate is thus dominated by the fastest eddy turnover time in the turbulence of each phase of the RTI.  In isotropic turbulence at high $Pm$, the growth is known to scale with the turnover time of Kolmogorov-scale eddies $\gamma \sim \delta v_{l_\nu}/l_\nu\sim (u/l_i)Re^{1/2}$, where $\delta v_{l}=(\epsilon l)^{1/3}$ is the velocity increment at length scale $l$,  $\epsilon\sim u^3/l_i$ is the turbulent kinetic energy dissipation rate, and the Kolmogorov length scale $l_\nu$ is defined by equating the momentum diffusion time and eddy turnover time $\nu/l_{\nu}^2\sim \delta v_{l_\nu}/l_\nu$ for an eddy of size $l\sim l_\nu$  \citep{Rincon2019}. This corresponds to ${D_{\gamma}}=1/2$ . In the following analysis, we restrict to the high $Pm$ limit and assume that the turbulence is sufficiently isotropic at viscous scales such that ${D_{\gamma}}=1/2$ in the mixing, saturation, and decay phases of the RTI. However, this assumption may need to be revisited in the mixing and decay phases, where the anisotropic effects of gravity are particularly important. Not only does gravity affect horizontal versus vertical fluid motions differently, there is also asymmetry between rising and falling fluid structures at increasingly higher Atwood numbers (for a review, see \citealt{zhou2017rayleigh}).

\paragraph{Model outline} In summary, we model the evolving velocity field driven by the RTI as isotropic turbulence with time dependent outer velocity scale $u(t)$ and integral scale $l_i(t)$ that drives the SSD with a time dependent dynamo growth rate $\gamma(t)= C_\gamma (u/l_i) Re^{\frac{1}{2}}$. The net amplification of magnetic energy, $\Delta$, can then be found by integrating $\gamma(t)$ in time across the duration of each phase of the RTI. We emphasize that we assume that the dynamo remains in the kinematic regime while the RTI evolves through the mixing, saturation, and decay phases, which is valid as long as the magnetic energy remains smaller than the kinetic energy at viscous scales.

A main goal of the paper is to determine how $\Delta$ scales with parameters of the RTI, $\{A,g,L,\nu\}$. We begin now by splitting up the time integral in Eq. \ref{MAF} into the three consecutive, turbulent phases and consider the positive contribution of each phase one at a time: 
\begin{equation}\label{DeltaPhases}
\Delta=\Delta^{mix}+\Delta^{sat}+\Delta^{decay}.
\end{equation}
\paragraph{RTI  Mixing phase} In the mixing phase, a mixing region of vertical extent $h(t)$ propagates away from the original interface. A variety of models have been proposed to predict the time dependence of $h(t)$, and the results of simulations and experiments are on various levels of disagreement that are primarily attributed to a sensitivity to initial conditions, system size effects, or effects of diffusivity. The general form of most models predict a freefall scaling given by 
\begin{equation}\label{height_ff}
    h(t)/L=\frac{1}{2}\alpha \tau^2+2 \sqrt{\alpha h_0/L}\tau+h_0/L,
\end{equation} 
where $\tau=t/t_{dyn}$, $\alpha$ is a constant, and $h_0$ is the length scale near when the mixing region first became nonlinear (see the review by \cite{Boffetta_2017} and references within). The turbulent velocity and speed of the mixing region boundary are taken to be comparable and self-consistently given by  $u(t)/u_{sat}=\alpha \tau+2 \sqrt{\alpha h_0/L}$. These scalings have been found to break down when the size of the bubbles and spikes become comparable to the horizontal or vertical extent of the domain or if there is a presence of a dominant mode $l_D$, both of which could lead to terminal velocity scalings $h(t)\sim \sqrt{g l_D}t$ and $u(t)\sim\sqrt{g l_D}$ \citep{dimonte2004dependence,BANERJEE20093906, Lecoanet2012}. Additionally, a larger, absolute thermal diffusivity reduces buoyancy effects and therefore slows the development of the mixing region \citep{Abarzhi_review}. For instance, dimensional arguments including diffusivity at low Atwood number can predict an alternative time dependence $h(t)\sim gt^2/\ln(gt^2/h_0)$ \citep{Abarzhi_2005}.

To model the dynamo growth rate, we assume that the turbulent integral scale is comparable to the height of the mixing region $l_i(t)\approx h(t)$ \citep{Chertkov_2003,Boffetta_2017}. However, this is an assumption that may also need to be revisited because the turbulence is not in a steady state and there may be a delay between energy generation at large scales and dissipation at small scales \citep{Livescu_2009}. For simplicity, we also assume the freefall scalings hold. Then the dynamo exponentially grows with rate:

\begin{equation}
    \gamma_{mix}(t)\approx \frac{C_\gamma Re_{sat}^{\frac{1}{2}}}{t_{dyn}}\left(\frac{(\alpha\tau+2 \sqrt{\alpha h_0/L})^3}{\frac{1}{2}\alpha \tau^2+2 \sqrt{\alpha h_0/L}\tau+h_0/L}\right)^{\frac{1}{2}}
\end{equation}
After a transient time $\tau\approx \sqrt{ h_0/(\alpha L)}$, the growth will asymptotically scale as $\gamma_{mix}(t)\sim t_{dyn}^{-1}Re_{sat}^{\frac{1}{2}}\tau^{\frac{1}{2}}$. Depending on the geometry, the mixing phase will last for a time interval $\Delta t_{mix}\sim t_{dyn}$ when the mixing region reaches the vertical system scale $h(t)\approx L$. Thus, with freefall scalings, $\Delta^{mix}$ is asymptotically found to be
\begin{equation}
    \Delta^{mix}\approx \frac{2\sqrt{2}\alpha C_\gamma}{3}Re_{sat}^{\frac{1}{2}}\left(\frac{\Delta t_{mix}}{t_{dyn}}\right)^{\frac{3}{2}}
\end{equation}

\paragraph{ RTI  Saturation phase} In the saturation phase, the mixing region encompasses the entire domain $h(t) = L\approx l_i(t)$ and the turbulent kinetic energy is maximal. The velocity scale $u(t)\approx \sqrt{2\alpha}u_{sat}$ can be solved either from the freefall scaling in the mixing phase or by equating converted initial potential energy $PE\sim  (\rho_t-\rho_n)\alpha g L^4$ to turbulent kinetic energy $KE\sim \frac{1}{2}(\rho_t+\rho_b)u^2L^3$. The dynamo growth rate is then $\gamma_{sat}(t)\approx C_\gamma (2\alpha)^{\frac{3}{4}} t_{dyn}^{-1}Re_{sat}^{\frac{1}{2}}$. We expect the velocity scale and integral length scale to remain roughly constant for a time interval $\Delta t_{sat}\sim t_{dyn}$ comparable to the dynamical time before the turbulence enters the decay phase. The exponential amplification from this phase should scale as 
\begin{equation}
    \Delta^{sat} \approx C_\gamma (2\alpha)^{\frac{3}{4}} Re_{sat}^{\frac{1}{2}}\left( \frac{\Delta t_{sat}}{t_{dyn}}\right)
\end{equation}

\paragraph{RTI Decay phase} From this point, the turbulence decays and the dynamo growth rate decreases. Again for simplicity, we assume that the free decaying turbulence is isotropic. Although the isotropy assumption is not expected to hold as the system relaxes into a stably stratified state, the simple model will be useful as a point of comparison. The total energy in isotropic turbulence decays as 

\begin{equation}
    \frac{dE(t')}{dt'}=-\frac{1}{2}\overline{\rho}C_E\frac{u(t')^3}{l_i(t')}
\end{equation}
where $u(t')=(2E(t')/\overline{\rho})^{\frac{1}{2}}$, $\overline{\rho}$ is the average density, $t'=0$ is the beginning of the decay phase, and $C_E$ is a dimensionless constant \citep{frisch1995turbulence,subramanian2006evolving}. The problem lies in determining the relation between the evolving integral scale and energy ($l_i(t')\sim E^{-s}$). Typical choices for $s$ lie in the range $\frac{1}{5}\leq s\leq\frac{1}{3}$ with $s=1/5$ corresponding to Batchelor-type and $s=1/3$ to Saffman-type turbulence, depending on properties and initial conditions of the turbulence \citep{ishida2006decay}. However, if the integral scale is already at the system scale and the system is bounded ( i.e. in a numerical simulation), then $l_i$ cannot grow and thus remains at the system scale ($l_i(t')\approx L$) independent of E (corresponding to $s\rightarrow 0$) \citep{skrbek2000decay,touil2002decay}. Solving this system for $u(t')$ and $l_i(t')$, the growth rate is given by:
\begin{equation}
    \gamma_{dec}(t')\approx \frac{C_\gamma(2\alpha)^{\frac{3}{4}} Re_{sat}^{\frac{1}{2}}}{t_{dyn}}\Big(1+(s+\frac{1}{2})C_E \tau'\Big)^{\frac{-3-2s}{2(1+2s)}}    
\end{equation}
The dynamo growth rate will eventually become zero when $Re(t')\approx Re^c(Pm)$. Assuming the dynamo is active in the decay phase from $\tau'=0$ to $\tau'=\Delta t_{dec}/t_{dyn}\gg 1/C_E$, $\Delta^{decay}$ is given by:
\begin{equation}
    \Delta^{decay}\approx  Re_{sat}^{\frac{1}{2}}\frac{4(2\alpha)^{\frac{3}{4}}C_\gamma }{C_E(1-2s)}
\end{equation}

\paragraph{Result} Combining our results,  each phase in Eq \ref{DeltaPhases} contributes  a term proportional to $Re_{sat}^{\frac{1}{2}}$ giving

\begin{align}
    \Delta\approx Re_{sat}^{\frac{1}{2}}C_\gamma \Bigg(&\frac{2\sqrt{2}\alpha}{3}\left(\frac{\Delta t_{mix}}{t_{dyn}}\right)^{\frac{3}{2}} \nonumber\\&+ (2\alpha)^{\frac{3}{4}}\frac{\Delta t_{sat}}{t_{dyn}}+\frac{4(2\alpha)^{\frac{3}{4}}C_\gamma }{C_E(1-2s)}\Bigg),
\end{align}
Since both $\Delta t_{mix}\sim t_{dyn}$ from freefall scaling and $\Delta t_{sat}\sim t_{dyn}$ from dimensional constraints, all the terms in the parenthesis are constants and thus:
\begin{equation}
    \Delta\sim Re_{sat}^{\frac{1}{2}}\sim\frac{A^{\frac{1}{4}}g^{\frac{1}{4}}L^{\frac{3}{4}}}{\nu^{\frac{1}{2}}},
\end{equation}
independent of constants in the models.

In summary, the clean result comes from the choice of freefall velocity ($u\sim u_{sat}$) and length ($l_i\sim L$) scales leading to a dynamo with growth rate $\gamma\sim t_{dyn}^{-1}Re_{sat}^{\frac{1}{2}}$ acting over dynamical timescales $t_{dyn}$ and resulting in amplification of $\Delta\sim \gamma t_{dyn}\sim Re_{sat}^{\frac{1}{2}}$.

\paragraph{Low $Pm$ case}
The RTI is also applicable to astrophysical scenarios where the magnetic Prandtl number $Pm=Rm/Re$ can be less than one, such as in stellar interiors during supernova explosions. The only modification is the estimate of the growth rate. The resistive scale at low $Pm$ is inside the inertial range ($l_\eta\sim Rm^{-3/4}L>l_\nu$) and the dynamo is thought to be driven by resistive-scale eddies whose turnover times are $t_{\eta}\sim L/(URm^{\frac{1}{2}})$ \citep{Iskakov_lowpm}. For $Re\gg Re^c(Pm)$, where $Re^c(Pm\ll 1)=\mathcal{O}(10^2)$, this corresponds to a reduced growth rate given by $\gamma\sim URm^{\frac{1}{2}}/L$ and an exponential amplification factor that instead scales as

\begin{equation}
\Delta\sim Rm_{sat}^{\frac{1}{2}}\sim \frac{A^{\frac{1}{4}}g^{\frac{1}{4}}L^{\frac{3}{4}}}{\eta^{\frac{1}{2}}}.
\end{equation}

\paragraph{Saturation of the SSD} We discuss the extension of our model of the dynamo in the kinematic regime to the dynamical regime and saturation of the SSD at high $Pm$. Our results above are only applicable in the kinematic regime of the dynamo where Lorentz forces are unimportant and the induction equation is a linear function of the velocity field, which leads to exponential growth of magnetic energy. However, when the magnetic energy becomes comparable to the kinetic energy at viscous scales ($\langle|\textbf{B}|^2 \rangle\sim Re^{-\frac{1}{2}}\langle|\textbf{u}|^2 \rangle$), the dynamo enters the dynamical regime where feedback on fluid motions from Lorentz forces become important and the dynamo switches to polynomial-order growth (see \cite{Rincon2019} and references therein). The dynamical regime is also known as the nonlinear growth phase, for which there are several models in the literature. In one possible model, the growth rate decreases due to sequential suppression of dynamo-generating motions by the Lorentz forces starting from the viscous scales and ending at the forcing scales \citep{Schekochihin_2002}. Assuming the growth rate is set by the turn over time of the smallest, unquenched velocity scale and that the magnetic energy is in equipartition with all smaller velocity scales, one can easily show that the magnetic energy will grow linearly in time $ME(t)\approx \zeta \epsilon t$, where $\epsilon\sim u^3/l_i$ is the transfer rate of kinetic energy and $\zeta$ is a dimensionless constant. When the Lorentz forces become strong at the forcing scales, the dynamo will fully saturate in near-equipartition between the total magnetic and kinetic energy with a ratio $ME/KE=f=\mathcal{O}(10^{-1})$, where $f$ is a dimensionless constant . 

Thus, our results for $\Delta$ are an upper bound because the SSD will leave the kinematic regime during some phase of the RTI if $\Delta$ is too large. As a rough estimate, any sub-equipartition, seed magnetic field configuration with magnetic energy greater than $ME(t=0)\gtrsim Re_{sat}^{-\frac{1}{2}}KE_{sat} e^{-\Delta}$ will reach the dynamical regime at some point in the RTI evolution and possibly saturate before the end of the decay phase of the RTI, which would result in MHD turbulence in the remaining RTI evolution. We briefly study this regime in Section \ref{sec:SaturationSimulation}.

While in principle one can build on our kinematic dynamo model and include the nonlinear dynamo growth phase by using a time dependent $\epsilon(t)$, we do not pursue this idea in further detail because (1) the nonlinear growth phase in simulations is short and difficult to test and (2) it is unclear how the feedback from the strong Lorentz forces across a growing range of velocity scales will affect the time evolution of the RTI velocity field. However, we do note a useful estimate for the time it takes for the dynamo to saturate during the dynamical regime $\Delta t_{d.r.}$. If the magnetic energy at the start of the dynamical regime is $ME\sim Re^{-\frac{1}{2}}KE$, at saturation is $ME=f\cdot KE$, and the transfer rate of kinetic energy is roughly constant $\epsilon \sim KE/t_{dyn}$, then we simply find that $\Delta t_{d.r.}$ is comparable to the dynamical time $\Delta t_{d.r.}\sim t_{dyn}(f-Re^{-\frac{1}{2}})/\zeta$.

\section{Simulation Scaling Analysis}\label{sec:SimulationScalingAnalysis}
In this section, we use three-dimensional direct numerical simulations of the RTI to test the theoretical scaling predictions presented in Section \ref{sec:TheoreticalScalingAnalysis}. We show and discuss the results of parameter scans with the Atwood number, gravitational acceleration, and viscosity.

\subsection{Numerical Setup}
We use the Athena++ code \citep{Athena} to solve the visco-resistive MHD equations in a 3D Cartesian domain with a similar setup to previous works \citep{Stone2007}.  The horizontal $-L/2\leq x,y\leq L/2$ directions have periodic boundary conditions while the vertical direction $- L \leq z\leq L$ has reflecting boundary conditions, where $L=0.1$. All simulations use a resolution of $N_xN_yN_z=512^2\times1024$, RK3 for the timestepper, and HLLD for the Riemann solver \citep{miyoshi2005multi}.

The initial hydrodynamic conditions set the fluid density $\rho(z)=\rho_t$ in the top half of the domain, $\rho(z)=\rho_b$ in the lower half of the domain, and a seed vertical velocity perturbation given by:

\begin{equation}
u_z(t=0)=\Re \left[\sum_{n_x,n_y}\frac{\widetilde{a}_{n_x,n_y}}{\sqrt{n}}e^{i\frac{2\pi}{L}\left(n_xx+n_y y\right)}\cos\left(\frac{\pi z}{2L}\right)\right]
\end{equation}
where $\widetilde{a}_{n_x,n_y}$ is a random complex number, $n=\sqrt{n_x^2+n_y^2}$, and $0< n\leq n_{max}$ with $n_{max}=32$.
The magnetic field is initialized with an isotropic spectrum $M(k)\approx \mathrm{const.}$ in the wavenumber range $0< k \leq k_{max}=2\pi n_{max}/L$. The total initial magnetic energy $E_M(t=0)=\int M(k)dk$ is set to a dynamically insignificant value $E_M(t=0)\approx10^{-19}$ when studying the kinematic regime in Section \ref{SimulationScaling}, and set higher when studying of the saturation regime. Lastly, we fix the magnetic Prandtl number to $Pm=3$ and the thermal Prandtl number to $Pr=\nu/\kappa=1$ for all runs.

Asymptotic scaling laws of hydrodynamic quantities in the mixing phase of the RTI are known to be sensitive to initial conditions (IC) and the box aspect ratio \citep{dimonte2004dependence,Boffetta_2017}.  Thus, parameter scans with different choices of IC (e.g. varying $n_{max}$) could lead to slightly different scaling laws. It is not clear which choice of IC is generally most physically applicable, since the RTI setup is already quite idealized. We discuss throughout this article how our dynamo results may vary for choices of IC different from ours. One common alternative option is mode perturbations with wavelengths that go down to the grid scale, while another is restricting to a shell of modes in Fourier space \citep{dimonte2004dependence}. Our choice of IC was motivated to allow us to study the effect of changing viscosity on the dynamo while retaining a similar hydrodynamic RTI evolution by having our fastest-growing mode to always be fixed at $k_{max}<k_c$ for our chosen range of viscosities.

In order to expect our results to be generalizable, it is important to at least check that minor variations of ICs for a fixed set of parameters have a minimal effect on $\Delta$. We have found this to generally be the case for a fiducial set of parameters where we widely varied $n_{max}\geq 8$ for both the initial velocity and magnetic field spectra. The two exceptions we found are the edge cases of a single mode RTI ($n_{max}=1$) or a purely uniform weak initial magnetic field. The single mode RTI becomes nonlinear near the system scale, unlike the multi-mode RTI, leading to negligible contribution from $\Delta^{mix}$ and a reduced $\Delta$. In the uniform magnetic field case, the dynamo additionally has a short and intense transient growth at the beginning of the mixing phase, due to coherent alignment of the field with the secondary shear layers that undergo the Kelvin-Helmholtz instability. We expect neither edge case to be applicable in a realistic astrophysical system.

\begin{figure}
    \centering
    \includegraphics[width=\linewidth]{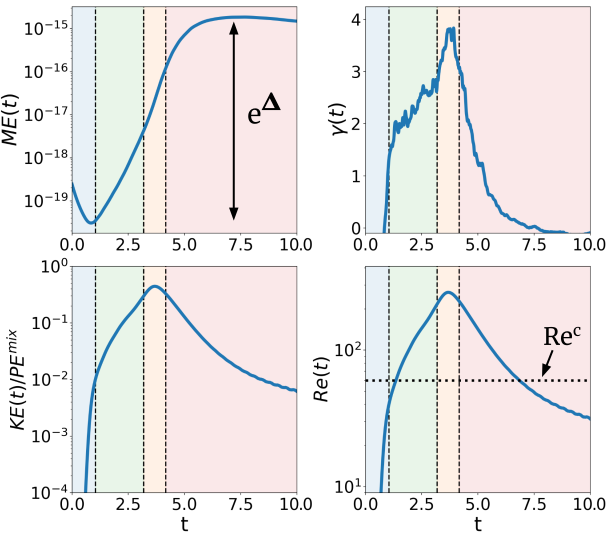}
    \caption{Evolution of the magnetic energy (top left), instantaneous dynamo growth rate (top right), kinetic energy (bottom left), and approximate Reynolds number (bottom right) of the fiducial simulation. The background shadings separated by vertical dashed lines denote the linear (blue), mixing (green), saturation (orange), and decay (red) phases. }
    \label{fig:fiducialSimPlot}
\end{figure}

The parameter scans are based on a fiducial simulation with parameters $\{A,g,\nu^{-1}\}=\{0.67,0.65,3\times 10^5\}$ which we use below to describe a typical simulation and explain our method of analysis. Figure \ref{fig:fiducialSimPlot} shows the time evolution of the kinetic energy, magnetic energy, dynamo growth rate, and approximate Reynolds number of the fiducial simulation. When the Reynolds number rises above the critical Reynolds number $Re(t)\gtrsim Re^c(Pm>1)\approx 60$, the dynamo growth rate sharply increases. The dynamo is triggered near the beginning of the mixing phase (green shaded region in Figure \ref{fig:fiducialSimPlot}), which we define as the time when $KE(t)=0.01 PE^{mix}$, where $PE^{mix}=0.5gL^4(\rho_t-\rho_b)$ is the potential energy released from complete mixing of the fluids. The kinetic energy rises until a maximum in the middle of the saturation phase at the saturation time $t^{(n)}_{sat}$ given by $KE(t^{(n)}_{sat})=\max(KE(t))$, at which point we also have the numerical values for $u^{(n)}_{sat}=u(t^{(n)}_{sat})$ and $t^{(n)}_{dyn}=L/u^{(n)}_{sat}$.

We define the mixing phase to end and the saturation phase (orange shaded region) to begin at time $t=t^{(n)}_{sat}-t^{(n)}_{dyn}$. Similarly, the saturation phase we define to end and the decay phase (red shaded region) to begin at time $t=t^{(n)}_{sat}+t^{(n)}_{dyn}$. At some point in the decay phase, the magnetic energy will reach its maximum and we define the decay phase to end for the purposes of the dynamo. Note that, while the exact definitions of the start and end time of each phase are somewhat arbitrary, we find that our results are not sensitive to reasonable variations of the definitions.

\begin{figure}
    \centering
    \includegraphics[width=\linewidth]{Slices.png}
    \caption{ Two-dimensional slices in the $X-Z$ plane of the density (top row) and vertical magnetic field (bottom row) in the mixing phase (left column), saturation phase (middle column), and decay phase (right column) of the fiducial simulation.}
    \label{fig:fiducialSlices}
\end{figure}

With the above definitions, we obtain the numerical values of the mean dynamo growth rate and amplification factor in each phase ($\overline{\gamma}_{phase}$ and $\Delta^{phase}$) with the following expressions:

\begin{equation}
\overline{\gamma}_{phase}=\frac{1}{\Delta t_{phase}}\int_{t_{beg}}^{t_{end}} \gamma(t)dt,
\end{equation}

\begin{equation}
\Delta^{phase}=\ln\left(\frac{ME(t_{end})}{ME(t_{beg})}\right)=\overline{\gamma}_{phase}\Delta t_{phase},
\end{equation} 

\begin{equation}
\Delta t_{phase}=t_{end}-t_{beg}
\end{equation}
where $t_{beg}$ and $t_{end}$ are the beginning and ending times of each phase. The total amplification factor is then very closely given by:
\begin{equation}
\Delta=\Delta^{mix}+\Delta^{sat}+\Delta^{dec}\approx\ln\left(\frac{\max(ME(t))}{\min(ME(t))}\right).
\end{equation}

For completeness, two-dimensional slices from the fiducial simulation of the density $\rho$ and vertical magnetic field $B^z$ during each of the three turbulent phases are shown in Figure \ref{fig:fiducialSlices}. The amplified magnetic field in the mixing phase (left column) closely follows the rising bubbles and falling spike structures at intermediate scales visible in the density field. The saturation phase (middle column) has large eddies on the system scale with the magnetic field developing on small scales, as would be qualitatively expected of isotropic turbulence. Last, the decay phase (right column) shows a negative mean density gradient in which residual kinetic energy sloshes fluid around while the magnetic fields still appear on scales that are small but slightly larger than in the saturation phase. 

\subsection{Scaling Results}\label{SimulationScaling}

\begin{figure*}
    \centering
    \includegraphics[width=\linewidth]{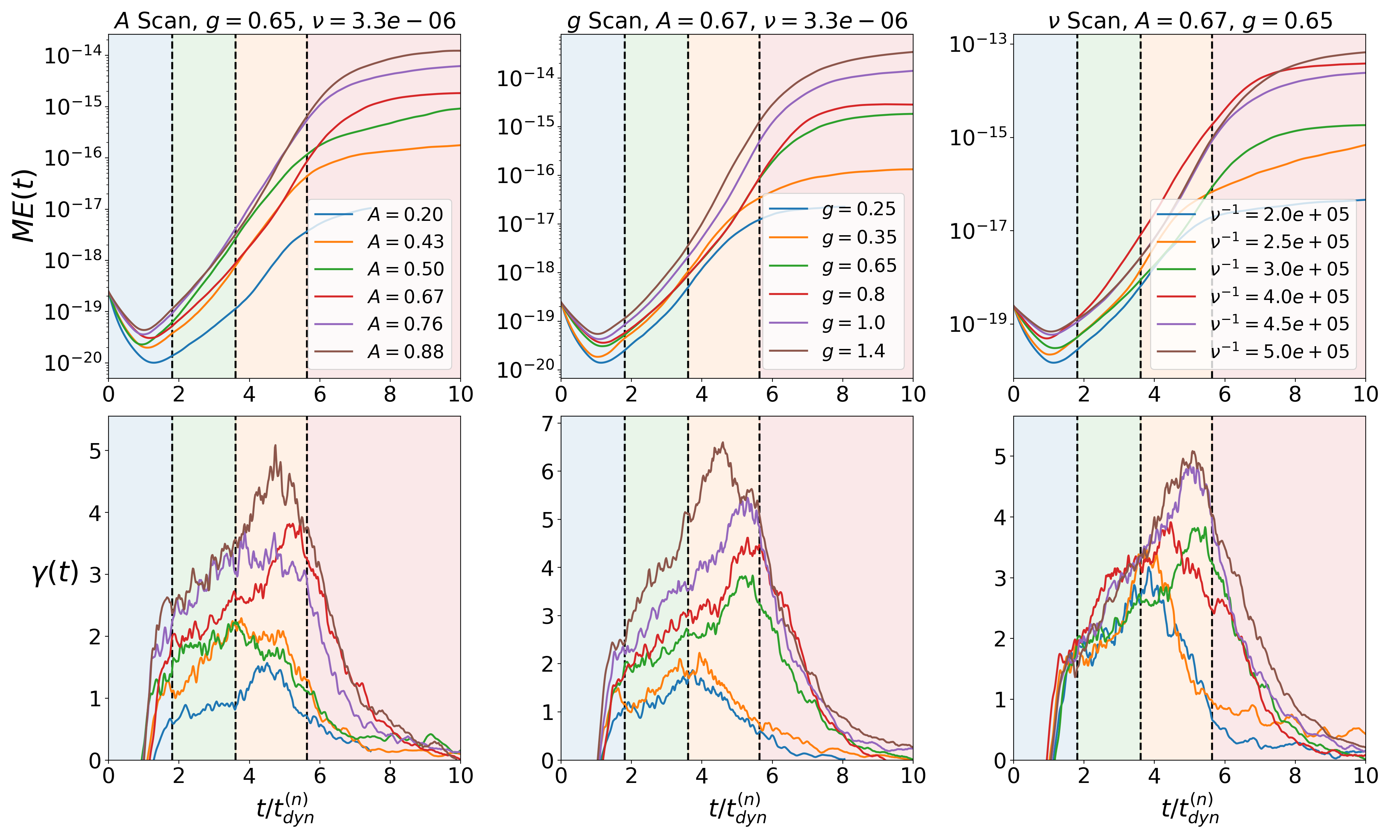}
    \caption{The magnetic energy (top row), $ME(t)$, and associated dynamo growth rate (bottom row) , $\gamma(t)$, versus time for the parameter scans with Atwood number (left column), gravitational acceleration (middle column), and viscosity (right column). Only six representative simulations from each parameter scan are shown for clarity of presentation. Legends of the figure in the top row also apply to the figures directly below.  The background shadings separated by dashed vertical lines are based on the mean duration of each phase of RTI (averaged over all simulations) and denote the linear (blue), mixing (green), saturation (orange), and decay (red) phases.  }
    \label{fig:ParameterScan}
\end{figure*}
To test the scaling relations of the dynamo model proposed in Section \ref{sec:TheoreticalScalingAnalysis}, we perform a series of parameter scans of the Atwood number, gravitational acceleration, and viscosity. We choose our fiducial simulation with values of $\{A,g,\nu^{-1}\}=\{0.67,0.65,3e5\}$ (corresponding to $\max(Re(t))\approx 300$) and then run a parameter scan across a maximum range of each parameter, while keeping all others fixed. For the gravitation acceleration parameter, we scan $g\in\{0.25,0.35,0.5,0.65,0.8,1.0,1.2,1.4\}$, for the Atwood number, we scan $A\in\{0.2,0.33,0.43,0.5,0.6, 0.67,0.71,0.76,0.82,0.88,0.9\}$, and for the viscosity, we scan $\nu^{-1}\in\{2,2.5,3,3.5,4,4.5,5\}\times 10^5$. Time evolution of the magnetic energies and dynamo growth rates for representative subsets of each parameter scan are shown in Figure \ref{fig:ParameterScan}. The lower bound for each parameter is constrained by needing $Re_{sat}\gg Re^c$ in order for the dynamo growth rate to be approximated by its asymptotic power law $\gamma \sim Re^{\frac{1}{2}}$, while the upper bound is constrained by requiring the Kolmogorov scale to be larger than the grid scale. In particular, the lowest value of the explicit viscosity is chosen to be a factor of two above the numerical viscosity for the numerical setup, which is estimated based on the decay rate of Alfven waves as described in Appendix \ref{sec:AlfvenWaveDecay}. Note that it is critical to at least marginally resolve the Kolmogorov scale, because otherwise the arbitrary simulation grid scale introduces another length scale that breaks the scaling arguments.

\begin{figure}
    \centering
    \includegraphics[width=\linewidth]{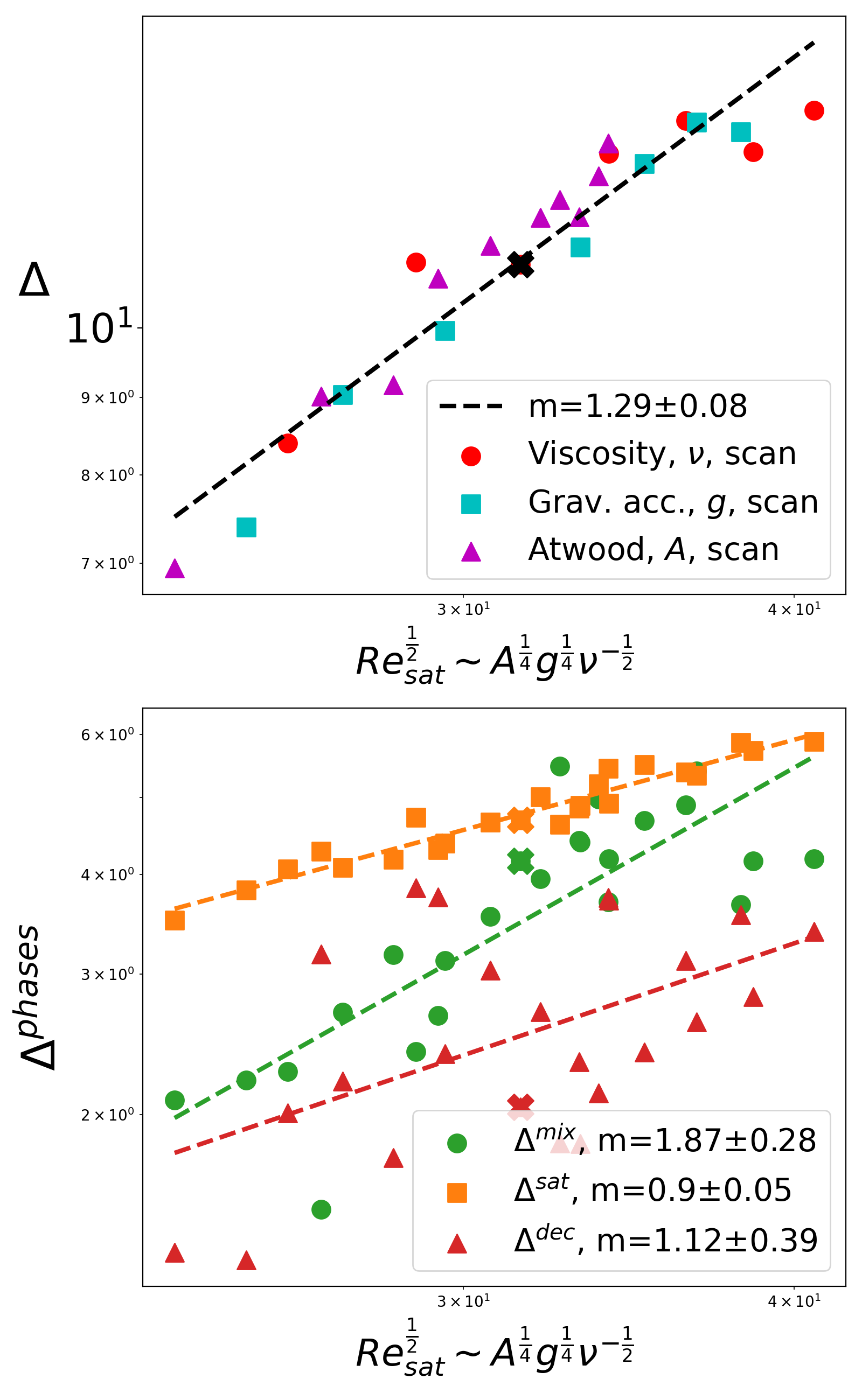}
    \caption{Top: The total magnetic energy amplification factor $\Delta$ versus $Re_{sat}^{\frac{1}{2}}=A^{0.25}g^{0.25}L^{0.75}/(2\pi\nu^{0.5})$ for each simulation on a $\log-\log$ scale is shown along with a linear fit $\ln \Delta=m\ln Re_{sat}^{\frac{1}{2}}+b$. The parameter scans are represented with red circles for the viscosity scan, cyan squares for the gravitational acceleration scan, and magenta triangles for the Atwood scan. Bottom: The total magnetic energy amplification factor $\Delta$ is split up into the contributions from the mixing (green circles), saturation (orange squares), and decay phases (red triangles) and plotted versus $Re_{sat}^{\frac{1}{2}}$.}
    \label{fig:DeltaScaling}
\end{figure}

First, we examine the dynamo amplification factor across the entire RTI instability. A linear fit in log-log space between $\Delta$ and each RTI parameter gives:  
 \begin{equation}
    \Delta\sim A^{0.40\pm 0.03} g^{0.35\pm 0.03}  \nu^{-0.51\pm 0.09},
\end{equation}
For reference, the predicted freefall scaling relations are given by:
 \begin{equation}
    \Delta\sim A^{0.25} g^{0.25}\nu^{-0.5}.
\end{equation}

While the viscosity exponent is in good agreement with the model (within one standard deviation), the exponents of $A$ and $g$ are close in magnitude to the predicted value of $0.25$, but are different by a statistically significantly amount (roughly five and three standard deviations). An alternative way to view the result is shown in the top panel of Figure \ref{fig:DeltaScaling} where the $\Delta$ is plotted versus $Re_{sat}^{\frac{1}{2}}$ on a log-log scale, so a linear fit $\ln \Delta=m\ln Re_{sat}^{\frac{1}{2}}+b$ with a slope of $m=1$ means perfect agreement. The data do appear to qualitatively follow a power law; however, the fit also shows a minor but statistically significant disagreement with the model quantified by the measured value of $m=1.29\pm 0.08$. In other words, with a measured value of $b=-0.90$, the simulations have a scaling $\Delta\approx 0.4Re_{sat}^{0.65\pm0.04}$ instead of the predicted $\Delta\sim Re_{sat}^{0.5}$.  Breaking up $\Delta$ further into the contribution of each phase $\Delta^{phase}$ in the bottom panel of Figure \ref{fig:DeltaScaling} clearly shows that the mixing phase is in disagreement while the saturation and decay phases are in good agreement with the model (although the decay phase data are highly scattered). 

To better understand the discrepancy, it is necessary to examine the scaling relations of the mean growth rate and duration of each phase with the RTI parameters. Following a linear fit in log-log space between $\{\Delta^{phase},\overline{\gamma}_{phase},\Delta t_{phase}\}$ and $\{A,g,\nu\}$ in each phase separately, Table \ref{tab:ScalingByPhase} shows the resulting scaling exponents. Overall, the freefall model predictions are again in good agreement for the saturation phase (all roughly within two standard deviations), while they show several disagreements for the mixing and decay phases. The deviation of the $m=1.29\pm 0.08$ result for $\Delta$ from the model prediction ($m=1$) can thus be explained by a likely failure of model assumptions in the mixing and decay phases. The value of $m=1.29$ is still close to $1$, however, because $\Delta_{sat}$ provides the dominant contribution to $\Delta$. We now present a more detailed analysis of the mixing and decay phases.

\begin{table}
    \centering
    \begin{tabular}{c|c|c|c}
         & $A$& $g$& $\nu$  \\
        \hline
         $\Delta_{ff}$ & 0.25& 0.25& -0.5  \\
         
         $\Delta_{mix}$ & 0.68$\pm$0.16& 0.43$\pm$0.10& -0.79$\pm$0.23  \\
         $\Delta_{sat}$ & 0.25$\pm$0.03& 0.24$\pm$0.02 & -0.38$\pm$0.04  \\
         $\Delta_{dec}$ & 0.28$\pm$0.23& 0.37$\pm$0.13& -0.35$\pm$0.33  \\
         \hline
         $\gamma_{ff}$ & 0.75& 0.75& -0.5  \\
         $\overline{\gamma}_{mix}$ & 1.01$\pm$0.05& 0.77$\pm$0.03& -0.54$\pm$ 0.07  \\
         $\overline{\gamma}_{sat}$ & 0.81$\pm$0.07& 0.81$\pm$0.05& -0.60$\pm$0.10  \\
         $\overline{\gamma}_{dec}$ & 0.17$\pm$0.19& 1.0$\pm$0.27& -1.3$\pm$0.36  \\
         \hline
         $\Delta t_{ff}$ & -0.5& -0.5& 0  \\
         $\Delta t_{mix}$ & -0.33$\pm$0.12& -0.34$\pm$0.08& -0.24$\pm$0.17 \\
         $\Delta t_{sat}$ & -0.56$\pm$0.06& -0.57$\pm$0.04& 0.21$\pm$0.10 \\
         $\Delta t_{dec}$ & 0.1$\pm$0.29& -0.62$\pm$0.38& 0.96$\pm$0.41 
    \end{tabular}
    \caption{Table of the scaling exponents between the dynamo amplification factor, mean dynamo growth rate, and duration of each phase (rows) and the RTI parameters (columns). For example, the entry for $\Delta_{mix}$ and $A$ means $\Delta_{mix}\sim A^{0.68\pm0.16}$. For reference, $\Delta_{ff}$, $\gamma_{ff}$, $\Delta t_{ff}$ denote the prediction for the exponent based on the freefall model.}
    \label{tab:ScalingByPhase}
\end{table}

\begin{figure}
    \centering
    \includegraphics[width=\linewidth]{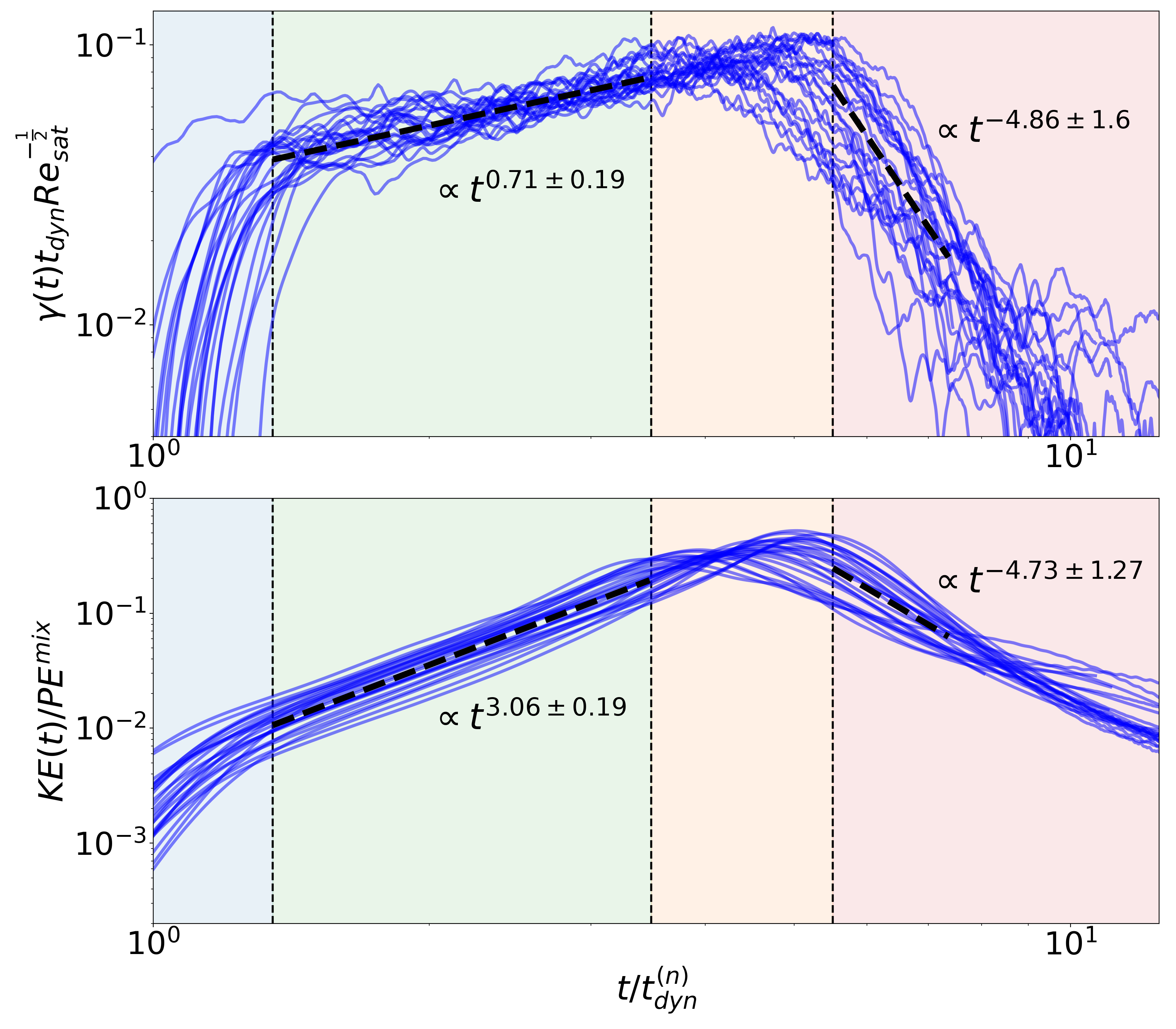}
    \caption{Time evolution of the dynamo growth rate (top) and kinetic energy (bottom) of every simulation. For each simulation, the time is rescaled by the dynamical time, $t_{dyn}$, the dynamo growth rate is rescaled by $t_{dyn}^{-1}Re_{sat}^{\frac{1}{2}}$, and the kinetic energy is rescaled by $PE^{mix}$.  The background shadings separated by dashed vertical lines are the same as in Figure \ref{fig:ParameterScan}. }
    \label{fig:ScalingWithTime}
\end{figure}

\begin{figure}
    \centering
    \includegraphics[width=\linewidth]{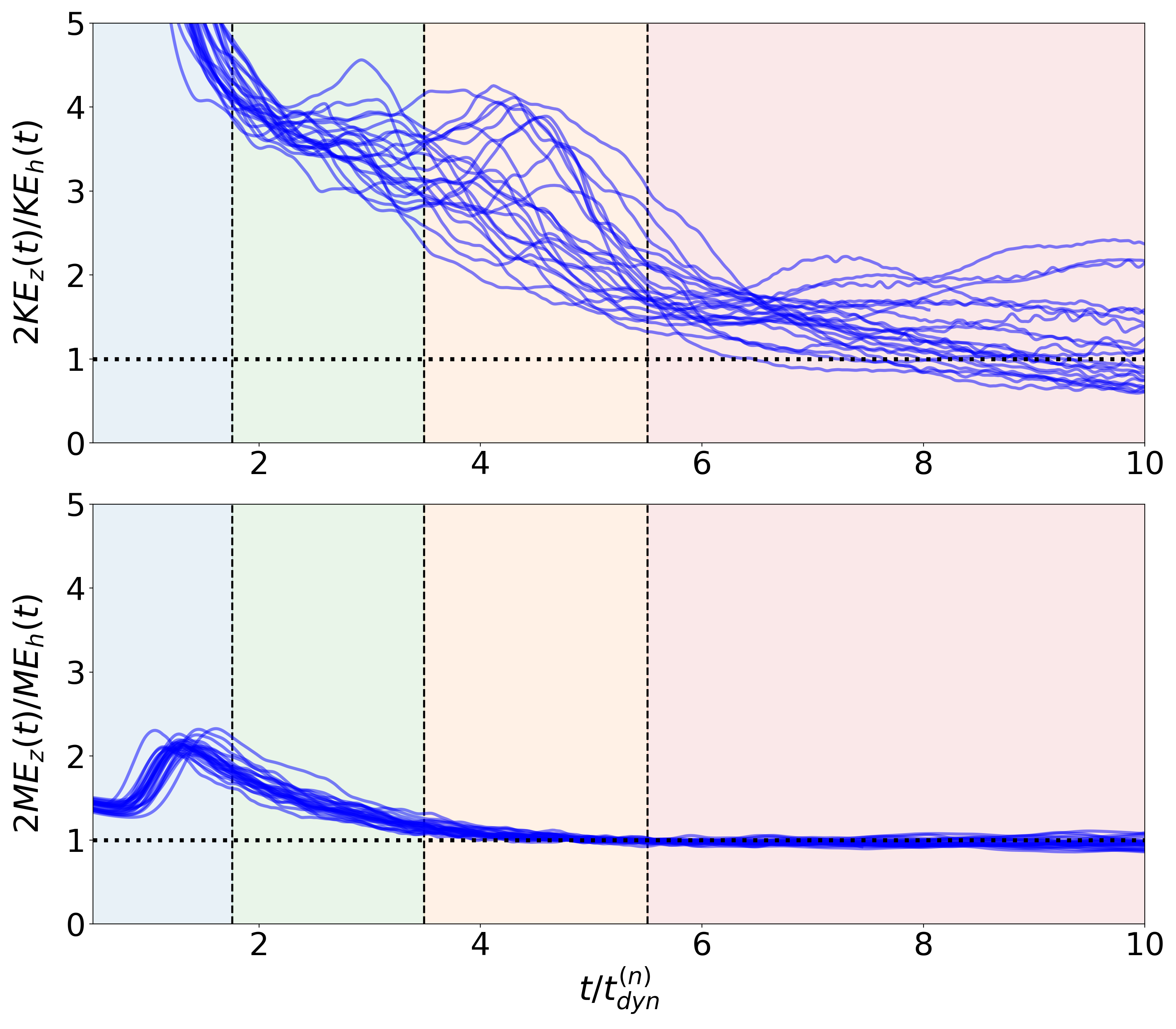}
    \caption{Time evolution of the anisotropy of total kinetic (top) and magnetic (bottom) energy of every simulation. The background shadings separated by dashed vertical lines are the same as in Figure \ref{fig:ParameterScan}.}
    \label{fig:Anisotropy}
\end{figure}

\paragraph{Mixing phase} The goal is to find which assumptions in the model of Section \ref{sec:TheoreticalScalingAnalysis} are violated in the mixing phase. We begin by comparing the time dependence of $KE(t)$ and $\gamma_{mix}(t)$ between freefall model predictions and rescaled simulation time series averaged over all runs, shown in Figure \ref{fig:ScalingWithTime}. The freefall model with the isotropy assumption asymptotically predicts a time dependence $\gamma_{mix}\sim \overline{u}(t)^{1.5}/l_i(t)^{0.5}\sim t^{0.5}$ for the dynamo growth rate and $KE(t)\sim \overline{u}(t)^2h(t)\sim t^4$ for the total kinetic energy with $ \overline{u}\sim t$ and $l_i(t)\sim h(t)\sim t^2$,
where
\begin{equation}
    \overline{u}(t)^2=\frac{\int_{-h(t)}^{h(t)}\int_{-L/2}^{L/2}\int_{-L/2}^{L/2} u(\textbf{x},t)^2d^3x}{2h(t)} 
\end{equation}

The numerical dynamo growth rate in the simulation data is slightly steeper, $\gamma^{(n)}_{mix}\sim t^{0.71\pm0.19}$ (within two standard deviations from the prediction), and the kinetic energy growth significantly shallower, $KE^{(n)}\sim t^{3.06\pm0.19}$ (four standard deviations from the prediction). The discrepancy suggests a slightly alternative scaling of either $\overline{u}(t)$, $h(t)$, or $l_i(t)$ in our simulations. Fitting the time power laws to $\overline{u}(t)$ and $h(t)$ independently for the fiducial simulation (not shown), we find the numerical time dependence $\overline{u}^{(n)}\sim t^{0.8\pm0.025}$ and $h^{(n)}\sim t^{1.55\pm0.06}$, which suggests an intermediate scaling between the freefall and terminal velocity scaling ($\overline{u}\sim t^0$, $h\sim t^1$) and has also been observed in previous simulations \citep{Lecoanet2012}. One possible reason we do not obtain freefall scaling may be our choice of initial conditions, which do not set $k_{max}=k_c$, while another reason may be that the vertical dimension of the box is not asymptotically large enough to reduce the effect of the linear term in Equation \ref{height_ff}. The effects of thermal diffusion may also play a role, despite the moderately high resolution of our simulations. The deviation of $\Delta t_{mix}$ from freefall predictions in Table \ref{tab:ScalingByPhase} is likely tied to the above reasons as well.

Nonetheless, the time dependence in the simulations predicts $\gamma_{mix}\sim {\overline{u}^{(n)}}^{1.5}/{h^{(n)}}^{0.5}\sim t^{0.45\pm0.05}$, which is even less steep than the freefall prediction. This motivates checking which of the remaining major assumptions break down: (1) isotropy of the dynamo-generating scales or (2) whether the integral scale and mixing height are directly proportional. The assumption of isotropy of turbulence at dynamo-generating scales appears to be supported by simulation data since ${D_{\gamma}}=d\ln \overline{\gamma}_{mix}/(d\ln \nu)\approx 0.5$ in Table \ref{tab:ScalingByPhase}. Additionally, while the  anisotropy of the total kinetic energy (defined by $2KE_z(t)/KE_h(t)$) is large in the mixing phase, as shown in the top panel of Figure \ref{fig:Anisotropy}, the anisotropy in the total magnetic energy (defined by $2ME_z(t)/ME_h(t)$) is much smaller and approaches unity by the end of the mixing phase (bottom panel of Figure \ref{fig:Anisotropy}). We expect the anisotropy of the magnetic field in the mixing phase to be even lower for larger Reynolds numbers because higher resolution simulations of the RTI than ours find strong evidence for isotropy at Kolmogorov scales \citep{Zingale_2005,Cabot2006}. 

We check the second assumption by computing $l_i(t)=(\int k^{-1}E(k,t) dk)/(\int E(k,t)dk)$ for the fiducial simulation and find $l_i^{(n)}\sim t^{1.1\pm0.04}$, which is statistically significantly different than the time dependence for $h^{(n)}\sim t^{1.55\pm0.06}$. Notably, the direct substitution $\gamma_{mix}\sim {\overline{u}^{(n)}}^{\frac{3}{2}}/{l_i^{(n)}}^{\frac{1}{2}}\sim t^{0.65\pm 0.04}$ has a better agreement with the observed $\gamma^{(n)}_{mix}(t)\sim t^{0.71\pm0.19}$. However, the large uncertainty of the time dependence of $\gamma^{(n)}_{mix}$ makes any conclusive determination difficult. 

Overall, a detailed understanding of the dynamo in the mixing phase of the RTI requires a further systematic study of the effects of initial conditions, box aspect ratio, and diffusivities, which we leave for future work. 

\begin{figure}
    \centering
    \includegraphics[width=\linewidth]{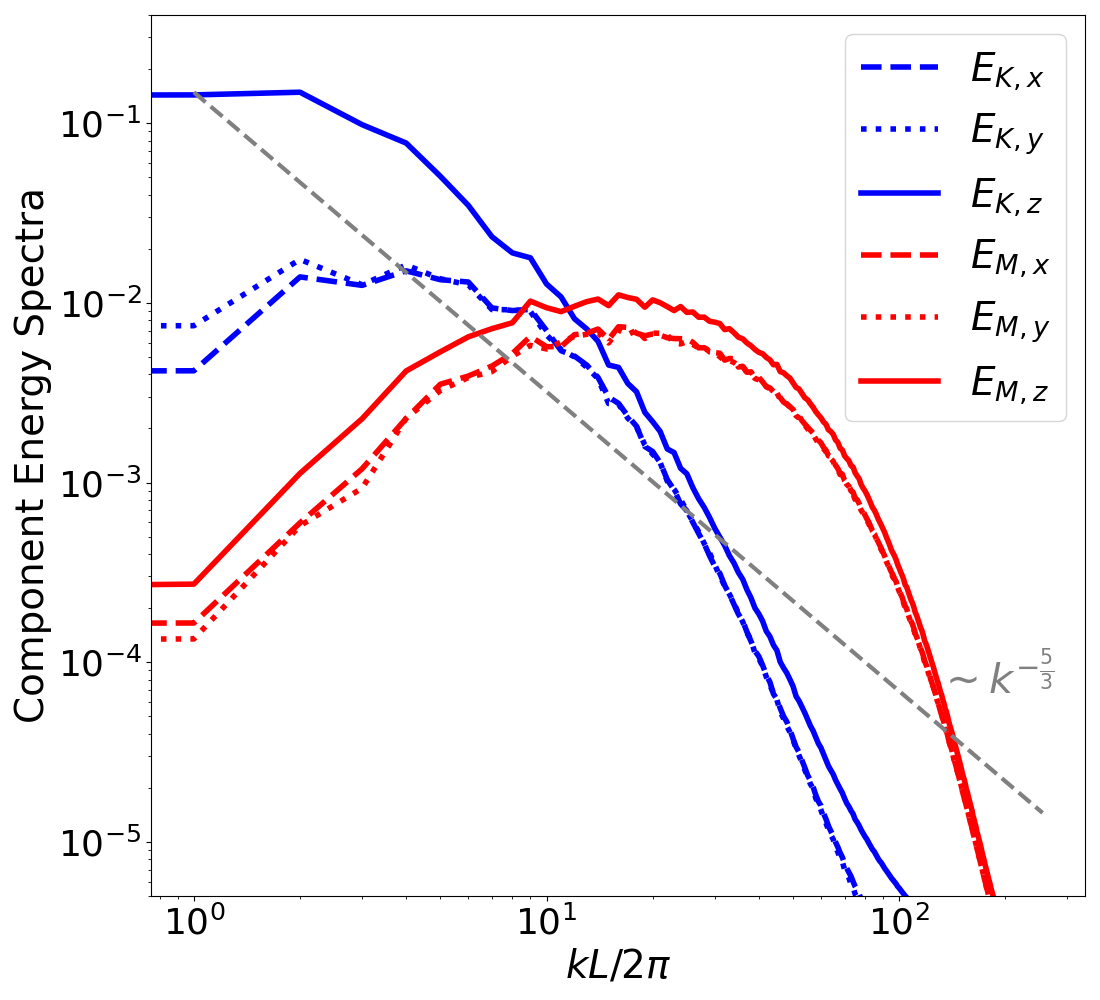}
    \caption{ Normalized component energy kinetic (blue), $E_{K,i}(k)$, and magnetic (red), $E_{M,i}(k)$, spectra in the mixing phase (at $t=2$) of the fiducial simulation. The kinetic and magnetic spectra are normalized by the total kinetic and magnetic energy at $t=2$ , respectively. }
    \label{fig:MixingPhaseSpectra}
\end{figure}

We speculate that there are two relatively important modifications to the model not examined in this study: (1) accounting for the non-steady-state nature of the mixing phase turbulence, and  (2) incorporating effects of large-scale anisotropy in the velocity field into the dynamo growth rate. The first modification requires including the time delay between the buoyancy forcing at the scale $l_i(t)$ and the dissipation rate $\epsilon(t)$, which is dissipating cascading energy due to forcing from earlier times \citep{Livescu_2009}. This is important because the dynamo operates at the dissipation scale and can more accurately be expressed as $\gamma(t)\sim (\epsilon(t)/\nu)^{\frac{1}{2}}$ \citep{Beresnyak_univSSD}, with $\epsilon(t)\sim u(t)^3/l_i(t)$ a good approximation only if the cascade rate is faster than the time rate of change of $l_i(t)$ and $u(t)$. 

The second modification may require incorporating the effect of large-scale velocity anisotropy in the dynamo growth rate. As shown in Figure \ref{fig:MixingPhaseSpectra}, the velocity spectra in the mixing phase are highly anisotropic at large scales with a dominant vertical component and become quasi-isotropic at intermediate and smaller scales (for $kL/2\pi\gtrsim 10$) while the magnetic field component spectra maintain roughly the same level of quasi-isotropy at all scales. Both of these patterns are observed in stably stratified turbulence, with the only difference being that the horizontal velocity components dominate at large scales instead \citep{Skoutnev_2021}. Thus, the effect of the large-scale anisotropy on the dynamo in the mixing phase can perhaps be modeled as a reduction in the effective Reynolds number through an effective Froude number, $Fr<1$, similar to the way the buoyancy Reynolds number, $Rb=Fr^2Re$, controls the dynamo growth rate in stably stratified turbulence \citep{Skoutnev_2021}. This extension may help explain the significant discrepancy of the exponent relating $\overline{\gamma}_{mix}$ and the Atwood number in Table \ref{tab:ScalingByPhase}, for instance.

\paragraph{Decay Phase}
The scatter in the data and uncertainty in the scaling exponents are unfortunately large in the decay phase. The uncertainty for some exponents in Table \ref{tab:ScalingByPhase} is comparable to the mean values, making it not possible to meaningfully compare with asymptotic predictions of the model. We attribute this to intermittency in the RTI turbulence, where occasionally a intermediate-scale, heavy parcel remains suspended for longer than usual and causes residual large-scale forcing, which can be seen as brief increases of $KE(t)$ and $2KE_z(t)/KE_h(t)$ in the decay phase of some runs in Figures \ref{fig:ScalingWithTime} and \ref{fig:Anisotropy}, respectively. The main assumption that the turbulence in the decay phase is freely decaying may perhaps be violated. It is not obvious whether residual forcing effects will persist in the decay phase at even higher Reynolds numbers. Additionally, the large uncertainty in $\overline{\gamma}_{dec}$ may also be a finite Reynolds number effect, since the rapid decay of $Re(t)$ from the moderate value of $Re$ in our simulations may violate the assumptions that $Re\gg Re^c$ and $\Delta t_{dec}\gg t_{dyn}/C_E$.

The asymptotic time dependence of the dynamo growth rate and kinetic energy similarly has large uncertainties. We find that $l_i(t)\approx t^0$ with $l_i\approx 0.3L$ throughout the entire decay phase, suggesting $s=0$ as expected since the turbulence has reached the box scale. This predicts an asymptotic scaling $\overline{u}(t)\sim t'^{-1}$, $KE(t) \sim t'^{-2}$, and $\gamma(t) \sim t'^{-1.5}$, where $t'=0$ is the beginning of the decay phase. The numerical power-law fit (Figure \ref{fig:ScalingWithTime}) gives  $KE^{(n)} \sim t'^{-4.73\pm1.27}$, and $\gamma^{(n)}\sim t'^{-4.86\pm1.6}$, which are both significantly steeper than the model prediction. The behavior of decaying RTI turbulence and associated dynamo is clearly poorly approximated by the model of freely decaying, isotropic turbulence. 

Fortunately, the decay phase has a smaller contribution to the total dynamo amplification factor (see bottom panel of Figure \ref{fig:DeltaScaling}) than either the mixing or saturation phases, which mitigates the effects of the large uncertainties. Reducing the uncertainties to better understand the decay phase will likely require simulations with much higher Reynolds numbers.

\subsection{Saturation}\label{sec:SaturationSimulation}
\begin{figure}
    \centering
    \includegraphics[width=\linewidth]{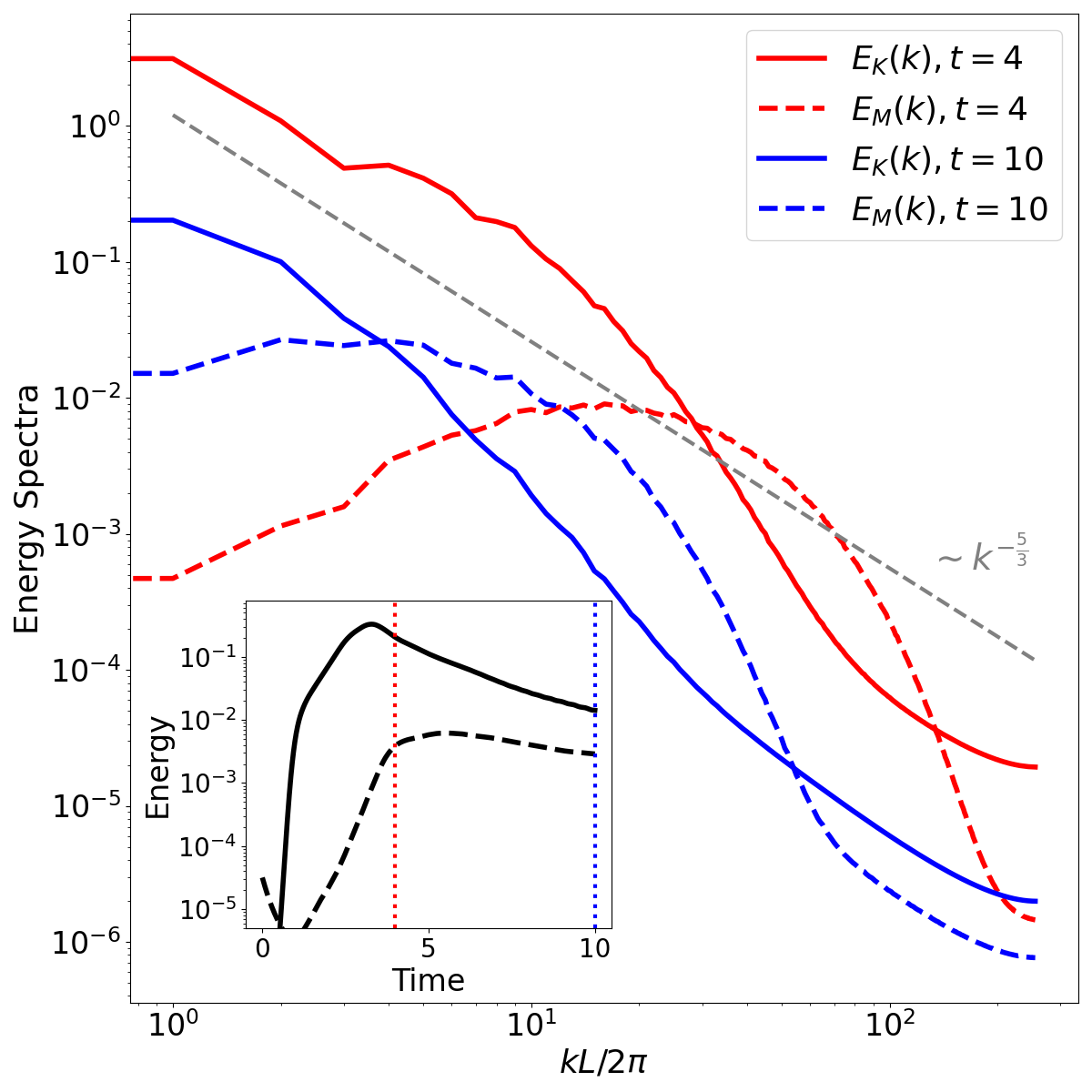}
    \caption{ Main figure shows the kinetic energy spectrum (solid lines) and magnetic energy spectrum (dashed lines) right after the saturation phase at $t=4$ (red) and in the decay phase (blue) at $t=10$. The spectra are normalized by the potential energy that would be released from complete mixing $PE^{mix}$. The inset shows the time evolution of the kinetic energy (solid lines) and magnetic energy (dashed lines), both also normalized by $PE^{mix}$. }
    \label{fig:SaturationSpectra}
\end{figure}

In this section, we study the case where the dynamo is able to saturate before the RTI fully relaxes. We simulate this regime by initializing the magnetic energy with $ME(t=0)\approx 10^{-5}PE^{mix}$ for the fiducial run, which satisfies the  $ME(t=0)\geq Re_{sat}^{-\frac{1}{2}}KE_{sat} e^{-\Delta}=\mathcal{O}(10^{-7})\cdot PE^{mix}$ criterion for the dynamo to reach the dynamical regime as discussed in Section \ref{sec:TheoreticalScalingAnalysis}. The energy evolution is shown in the inset of Figure \ref{fig:SaturationSpectra} where the dynamical regime is observed to begin around $t\approx 4$ when the slope of the magnetic energy growth sharply drops. The following nonlinear growth phase appears to last only briefly until $t\approx5$ ($t_{dyn}\approx0.5$ for the fiducial simulation).  This is expected because the $Re_{sat}^{-\frac{1}{2}}\sim 0.03$ for the fiducial simulation and the final ratio of magnetic to kinetic energy is $f\approx 0.1$, leading to a short predicted time to saturation in the dynamical regime of $\Delta t_{d.r.}\sim t_{dyn}(f-Re_{sat}^{-\frac{1}{2}})$.  

To qualitatively test the phenomenological model of the dynamo in the dynamical regime discussed in Section \ref{sec:TheoreticalScalingAnalysis}, we plot the isotropic energy spectra of a cubic volume in the region $-L/2\leq z\leq L/2$ at two representative times $t\in\{4,10\}$ in Figure \ref{fig:SaturationSpectra}. At $t=4$ (red curves) just after the dynamo has enter the nonlinear growth phase, the magnetic energy is larger on a scale-by-scale basis below an intermediate scale $kL/2\pi\sim30$. In the fully saturated phase of the dynamo at $t=10$ (blue curves), the magnetic energy is larger than the kinetic energy at basically all but the largest scales. Both of these observations are in accord with the predictions from the current understanding of the dynamical dynamo regime in steady-state turbulence as discussed in Section \ref{sec:TheoreticalScalingAnalysis}.

Overall, these results show that the RTI is capable of generating small-scale magnetic fields in near-equipartition ($ME/KE^{dec}=f =\mathcal{O}(10^{-1})$) with the decaying turbulent kinetic energy, $KE^{dec}$, which is a fraction on the order of $KE^{dec}/PE^{mix}=\mathcal{O}(10^{-2})$ of the initially available potential energy. These magnetic fields then can act as seed fields for further processes driven by larger scales in the astrophysical object. 
 
\section{Rayleigh-Taylor driven dynamo in neutron star mergers}\label{sec:Application}
One open question in recent studies of binary neutron star mergers is the source, efficiency, and time scale of magnetic field amplification in the post-merger star \citep{Price:2006fi,Kiuchi:2015sga,2015ApJ...809...39G}. Several studies have proposed the role of the Kelvin-Helmholtz instability (KHI) active at the contact region of the two stars in contributing to the magnetic field amplification observed in the core of the post-merger \citep{Kiuchi:2015sga,Aguilera-Miret:2020dhz}.  It is generally assumed that the Reynolds number of the KHI turbulence is high enough that the dynamo will saturate in near-equipartition with the kinetic turbulence, although it is unclear how fast that will happen \citep{2018PhRvD..97l4039K}. The KHI, however, cannot explain the large amount of dynamo action that is observed in the surface layer during the merger \citep{Kiuchi:2015sga}. We propose that the RTI is likely responsible for dynamo action in the outer parts of the merger where centrifuged denser matter can be seen falling down onto less dense matter at adjacent longitudes. Rapid saturation of the dynamo and the associated amplification of the magnetic field to near-equipartition with the RTI-driven turbulence might critically affect the prospects of long-term mass ejection \citep{Metzger_2018,Ciolfi:2020wfx}, as well as jet launching \citep{Ciolfi:2020hgg,Mosta:2020hlh} from a stable magnetar remnant. Using our model, we verify that the RTI-driven dynamo does saturate, and we provide a quantitative estimate for the time scale for saturation in the conditions of the post-merger neutron star envelope.

We first need an estimate for the Reynolds number $Re_{sat}$. The viscosity of neutron star matter is strongly dependent on the temperature and density conditions.
Assuming low temperatures of $T\simeq 1\,\rm MeV$ and densities around nuclear saturation, i.e. $2\times 10^{14}\, \rm g/cm^3$, the kinematic shear viscosity $\nu\sim {3\times 10^{-3}}\, \rm m^2/s$ in the electron-scattering-dominated regime \citep{Shternin:2008es}.
On the other hand, for temperatures $T\simeq 10\, \rm MeV$, neutrino-emitting Urca processes (e.g. $n\rightarrow p\, e^-\, \bar{\nu}_e$ and $p \, e^-\rightarrow n\, \nu_e$ ) might be the dominant contribution, leading to $\nu \simeq 10^4 \, \rm m^2/s$ at densities around saturation \citep{Alford:2017rxf}. The resistivity for a warm ($T\simeq 1\, \rm MeV$) neutron star crust is set by electron scattering of correlated nuclei \citep{Harutyunyan:2016rxm} and has a value of $\eta\simeq 5\times 10^{-7} \rm m^2/s$ \citep{Harutyunyan:2018mpe}, which puts neutron star matter well into the high $Pm$ regime with $Pm=\mathcal{O}(10^{4-11})$. The extreme density gradients at the surface of a neutron star, imply that the Atwood number $A\approx1$. We consider a thin layer exhibiting these gradients and becoming RTI-unstable on a length scale that is a fraction of a scale height $L\sim 0.1H$ ($H\sim 1\,\rm km$). We approximate the gravitational acceleration by means of the Newtonian expression for surface gravity, i.e. $g\simeq \frac{GM}{R^2}\approx 3\times 10^{16}\, \rm m/s^2$. Although $u_{sat}$ as obtained from the expressions above technically would be superluminal (owing to the simplified Newtonian assumptions made here), we take this as an indication that $u_{sat}\simeq c$. Finally, all these parameters correspond to an estimate $Re_{\rm sat}\sim \mathcal{O}(10^6-10^{13})$. If we assume our model holds ($\Delta\approx C_\Delta Re_{sat}^{\frac{1}{2}}$ with $C_\Delta=\mathcal{O}(1)$), then in a RTI-unstable region the expected amplification would be $\Delta\sim \mathcal{O}(10^{3}-10^{6})$. The assumption that the dynamo will reach equipartition with the kinetic energy of the viscous scales, $KE/Re^{\frac{1}{2}}$ is easily justified since $e^{\Delta}\gg KE/(ME(0)Re^{\frac{1}{2}})$ even for generous estimates of order $KE/ME(0)\sim 10^{\mathcal{O} (10)}$, where $ME(0)$ is the initial magnetic energy density near the surface of one of the initial neutron stars and $KE$ is the characteristic turbulent kinetic energy in post-merger envelope.

We can now estimate how long it will take the dynamo to leave the kinematic regime ($\Delta t_{k.r.}$) using the dynamo growth rate $\gamma\sim t_{dyn}^{-1}Re_{sat}^{\frac{1}{2}}\sim \mathcal{O}(10^{3}-10^{7}) \rm \mu s^{-1}$, where we have used $t_{dyn}= L/u_{\rm sat}\sim 0.3 \,\rm \mu s$ for the dynamical time. The estimate is as follows:
\begin{align}
    \Delta t_{k.r.}\sim& \frac{1}{\gamma}\ln \left(\frac{KE}{Re_{sat}^{\frac{1}{2}}ME(0)}\right)\nonumber \\
    \sim &
     2\log_{10} \left(\frac{KE}{Re_{sat}^{\frac{1}{2}}ME(0)}\right)\, \times \left[10^{-7} - 10^{-3}\right]\, \rm \mu s \nonumber \\ 
     \ll & t_{\rm dyn}.
\end{align}
After the kinematic regime, the dynamo will continue to grow in the dynamical regime and fully saturate on a timescale of
\begin{align}
    \Delta t_{d.r.}&\sim t_{dyn}(f-Re_{sat}^{-\frac{1}{2}})/\zeta \nonumber \\
    &\sim t_{dyn}f/\zeta\nonumber\\
    &\sim\mathcal{O}(10^{-1})\rm \mu s 
\end{align} 
using the model of the nonlinear dynamo growth phase described in Section \ref{sec:TheoreticalScalingAnalysis} and assuming $f=\mathcal{O}(10^{-1})\gg Re_{sat}^{-\frac{1}{2}}$.

Since the duration of the kinematic dynamo regime (upper bound of nanoseconds) and duration of the dynamical dynamo regime (upper bound of microseconds) of the RTI-driven turbulence in the envelope are both much smaller than the relaxation time of the merger (order of milliseconds), we expect magnetic energies to be in near-equipartition with the kinetic energy of the RTI-driven turbulence in the envelope across essentially the entire of duration the post-merger evolution.

\section{Summary and Conclusions}\label{sec:Conclusion}
We present a model for the kinematic small-scale dynamo in the mixing, saturation, and decay phases of the Rayleigh Taylor instability of an ionized, collisional plasma with freefall and isotropy assumptions for the turbulence in each of the phases. The model quantitatively predicts scaling relations between the properties of the dynamo (growth rate and total magnetic energy exponential amplification factor) and parameters of the RTI (Atwood number, gravitational acceleration, length scale, and viscosity). The model predictions are tested with sets of three-dimensional direct numerical simulations that solve the visco-resistive MHD equations using the Athena++ code. The main results are itemized below:

\begin{itemize}
    \item We find that the total magnetic energy exponential amplification factor, $\Delta$, based on simulation data scales as 
    \begin{equation}
        \Delta\approx C_\Delta Re_{sat}^{D_{\gamma}}
    \end{equation} with constants $C_\Delta\approx 0.4$ and ${D_{\gamma}}\approx0.65\pm0.04$. This is in fairly close agreement with the model prediction of ${D_{\gamma}}=0.5$, but the difference is statistically significant. An analysis of the dynamo in each phase reveals that the model correctly predicts scaling relations in the saturation phase, while having several discrepancies in the mixing and decay phases.
    \item An analysis of the dynamo scaling relations and time dependence in the mixing phase shows that the freefall and isotropy assumptions are fairly well-supported with a few minor discrepancies, which are attributed to deviations of the evolution of the hydrodynamic turbulence from freefall predictions. For example, we do not find strong agreement with the freefall predictions of quadratic scaling of mixing height with time, linear scaling of the root-mean-square velocity with time, nor constant proportionality between mixing height and the instantaneous integral scale. These discrepancies are well-known in the literature and are primarily attributed to the choice of initial conditions and the effects of finite diffusivities in simulations. We leave a more detailed analysis of the SSD in the mixing phase for future studies.
    \item In the decay phase, the dynamo scaling relations and time dependencies have large uncertainties and our model assumption of freely decaying isotropic turbulence is not a good fit. We attribute this to residual buoyant forcing and possibly finite Reynolds number effects. Fortunately, the magnetic amplification in the decay phase is small compared to the contributions from the mixing and saturation phases.
    \item In the saturation regime of the small-scale dynamo, the magnetic field is found to reach near-equipartition with the large scales and super-equipartition with the intermediate and small scales of the decaying velocity field of the RTI. We study this regime by running a single simulation with a moderate initial magnetic field so that the dynamo reaches saturation before the RTI fully relaxes.  
    \item We propose that the small-scale dynamo driven by RTI turbulence helps explain observations of magnetic energy amplification in the outer regions of the post-merger in global simulations, complementary to amplification by saturation of the the Kelvin-Helmholtz instability observed in the core. Applying the scaling relations to the parameter regime of RTI-unstable regions of the outer envelopes of binary neutron star mergers, the model predicts that the kinematic regime of the small-scale dynamo will end on the time scale of nanoseconds and then reach saturation on a timescale of microseconds, which are both fast in comparison to the millisecond relaxation timescale of the merger. 
\end{itemize}
The flexibility of the model allows for easy extensions in future studies of the dynamo in RTI turbulence. The model primarily prescribes a time dependence for the instantaneous integral length and outer velocity scale in each phase of the RTI, which are then substituted into an equation for the dynamo growth rate. Using alternative time dependencies based on the choice of initial conditions, or allowing a time delay between forcing and dissipation, could improve understanding of the SSD in the mixing phase, for example. We leave such extensions for future work.

\acknowledgments
We acknowledge the Flatiron’s Center for Computational Astrophysics (CCA) and the Princeton Plasma Physics Laboratory (PPPL) for the support of collaborative CCA-PPPL meetings on plasma-astrophysics where insightful comments and discussions contributed to this work. Research at the Flatiron Institute is supported by the Simons Foundation. Simulations were carried out on the Frontera cluster with NSF Frontera grant number AST20008. A. B. was supported by the DOE Grant for the Max Planck Princeton Center (MPPC). A.P. acknowledges support by the National Science Foundation under Grant No. AST-1909458. E.R.M. gratefully acknowledges support from a joint fellowship at the Princeton Center for Theoretical Science, the Princeton Gravity Initiative and the Institute for Advanced Study. V. S. was supported by Max-Planck/Princeton Center for Plasma Physics (NSF grant PHY-1804048).

\appendix
\begin{figure}
    \centering
    \includegraphics[width=0.5\linewidth]{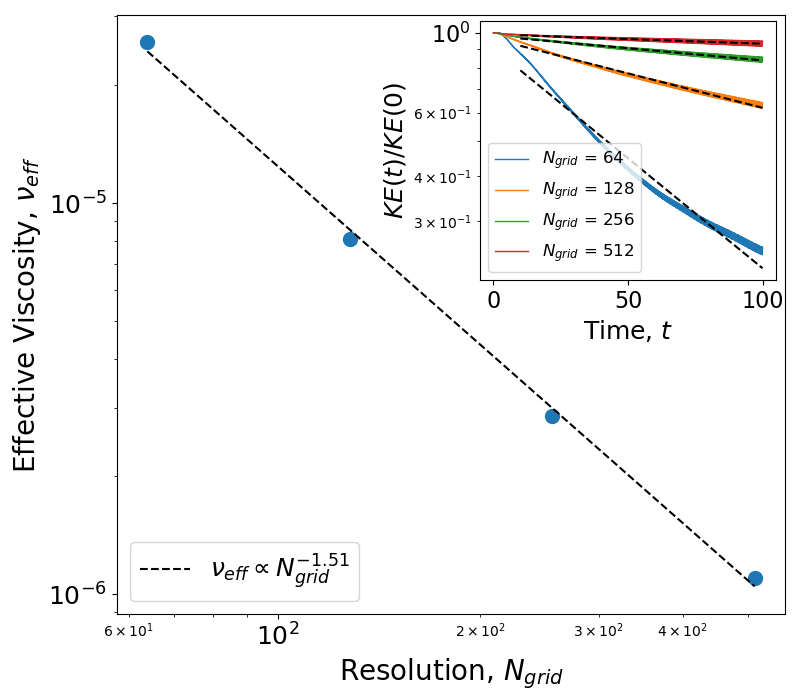}
    \caption{ Main figure shows the effective numerical viscosity vs. the grid resolution for our numerical setup. The inset plot shows the the exponentially decaying kinetic energy (solid lines) of the Alfven wave at different resolutions and fits (dashed black lines) that provide estimates for the effective numerical viscosities.  }
    \label{fig:EffectiveViscosity}
\end{figure}\section{Decaying Alfven Wave}\label{sec:AlfvenWaveDecay}
In a grid-based code like Athena++, numerical diffusivity acts as a resolution-dependent and algorithm-dependent effective viscosity that regularizes the turbulent cascade in a hydrodynamical simulation without an explicit viscosity. An explicit viscosity will only be meaningful if it is larger than the effective viscosity. We estimate the effective numerical viscosity of our numerical setup (RK3 for the timestepper and HLLD for the Riemann solver) by launching an Alfven wave along the main diagonal in a cubical domain with zero explicit viscosity and measuring the decay rate. The Alfven wave has wavenumber $\textbf{k}=(1,1,1)\cdot 2\pi/L$ in a background field $\textbf{B}_0=(1,\sqrt{2},0.5)$ and the domain has triply periodic boundary conditions with a resolution $N_{grid}^3$. The kinetic energy of the wave will decay approximately exponentially $KE(t)\sim e^{-2\nu_{eff}\textbf{k}^2t}$. We measure $\nu_{eff}$ with a linear fit on a plot of $\log (KE(t))$ versus time for four grid resolutions as shown in the inset of Figure \ref{fig:EffectiveViscosity}. The main plot in Figure \ref{fig:EffectiveViscosity} shows a clean power-law fit $\nu_{eff}\sim N_{grid}^{-1.51}$. The value of the numerical viscosity is approximately $\nu_{eff}\approx  10^{-6}$ for the resolution $N_{grid}=512$ used in our simulations of the RTI. This informs our choice of the explicit viscosity $\nu\approx 3\cdot 10^{-6}$ for the fiducial simulation and $\nu= 2\cdot 10^{-6}$ for our lowest choice of viscosity in the viscosity parameter scan in Section \ref{sec:SimulationScalingAnalysis}.


\vspace{5mm}

\software{Athena++ \citep{Athena}}


\bibliography{main}{}
\bibliographystyle{aasjournal}



\end{document}